\documentclass[conference]{IEEEtran}

\usepackage{amsmath}
\usepackage{graphicx}
\usepackage{enumerate}
\usepackage[noend]{algorithmic}
\usepackage[noend]{algorithm}
\usepackage{listings}
\usepackage{subfigure}

\def\includefigs{\let\ifincfigs=\iftrue}
\def\noincludefigs{\let\ifincfigs=\iffalse}
\includefigs

\newbox\epsfvertlab
\newbox\epsfhorlab
\newbox\epsffiglab

\newdimen\epsfvlabsize
\newdimen\scott

\def\setvlabel#1{\setbox\epsfvertlab=\vbox{\hbox{#1}}}%
\def\sethlabel#1{\setbox\epsfhorlab=\vbox{\hbox{#1}}}%
\def\figlab#1 #2 #3{\setbox\epsffiglab=\vbox to 0pt{%
\ifvoid\epsffiglab\else\box\epsffiglab\fi\vss\hbox to 0pt{\raise #2 \hbox{\hskip #1 #3}\hss}}}

\newdimen\fighor
\newdimen\figver
\newbox\rotbox
\long\def\lrlap#1{\hbox to 0pt{#1\hss}}
\long\def\verttex#1#2#3{{\fighor = #1\figver = #2\vbox to \figver{\vss%
\hbox to \fighor{\hfill\hsize=\fighor%
\lrlap{\rotstart{-90 rotate}\vbox to \fighor{#3\vfil}\rotfinish}}}}}

\def\dvipsvspec#1{\special{ps:#1}}
\def\dvipsrotstart#1{\dvipsvspec{gsave currentpoint currentpoint translate
   #1 neg exch neg exch translate}}
\def\dvipsrotfinish{\dvipsvspec{currentpoint grestore moveto}}

\def\rotstart#1{\dvipsrotstart{#1}}
\def\rotfinish{\dvipsrotfinish}

\def\epsfsetlab{%
\ifvoid\epsfvertlab%
\else%
\verttex{\epsfvlabsize}{\epsfysize}%
{\hbox to \epsfysize{\hss\box\epsfvertlab\hss}}%
\fi%
\ifvoid\epsfhorlab%
\else%
\scott=\epsfxsize%
\advance\scott by \epsfvlabsize%
\rlap{\vtop{\hrule height0pt\hbox to \scott{\hss\box\epsfhorlab\hss}}}%
\fi%
}

\def\epsfsetover{\ifvoid\epsffiglab\else\box\epsffiglab\fi}

\newread\epsffilein    
\newif\ifepsffileok    
\newif\ifepsfbbfound   
\newif\ifepsfverbose   
\newdimen\epsfxsize    
\newdimen\epsfysize    
\newdimen\epsftsize    
\newdimen\epsfrsize    
\newdimen\epsftmp      
\newdimen\pspoints     
\def\epsfbox#1{
   \ifvoid\epsfvertlab%
   \else\epsfvlabsize=\ht\epsfvertlab \advance\epsfvlabsize by \dp\epsfvertlab\fi%
   \leavevmode\global\def\epsfllx{72}\global\def\epsflly{72}%
   \global\def\epsfurx{540}\global\def\epsfury{720}%
   \def\lbracket{[}\def\testit{#1}\ifx\testit\lbracket
   \let\next=\epsfgetlitbb\else\let\next=\epsfnormal\fi\next{#1}}%
\def\epsfgetlitbb#1#2 #3 #4 #5]#6{\epsfgrab #2 #3 #4 #5 .\\%
   \epsfsetgraph{#6}}%
\def\epsfnormal#1{\epsfgetbb{#1}\epsfsetgraph{#1}}%
\def\epsfgetbb#1{%
\openin\epsffilein=#1
\ifeof\epsffilein\errmessage{I couldn't open #1, will ignore it}\else
   {\epsffileoktrue \chardef\other=12
    \def\do##1{\catcode`##1=\other}\dospecials \catcode`\ =10
    \loop
       \read\epsffilein to \epsffileline
       \ifeof\epsffilein\epsffileokfalse\else
          \expandafter\epsfaux\epsffileline:. \\%
       \fi
   \ifepsffileok\repeat
   \ifepsfbbfound\else
    \ifepsfverbose\message{No bounding box comment in #1; using defaults}\fi\fi
   }\closein\epsffilein\fi}%
\def\epsfsetgraph#1{%
   \epsfrsize=\epsfury\pspoints
   \advance\epsfrsize by-\epsflly\pspoints
   \epsftsize=\epsfurx\pspoints
   \advance\epsftsize by-\epsfllx\pspoints
   \epsfxsize\epsfsize\epsftsize\epsfrsize
   \ifnum\epsfxsize=0 \ifnum\epsfysize=0
      \epsfxsize=\epsftsize \epsfysize=\epsfrsize
     \else\epsftmp=\epsftsize \divide\epsftmp\epsfrsize
       \epsfxsize=\epsfysize \multiply\epsfxsize\epsftmp
       \multiply\epsftmp\epsfrsize \advance\epsftsize-\epsftmp
       \epsftmp=\epsfysize
       \loop \advance\epsftsize\epsftsize \divide\epsftmp 2
       \ifnum\epsftmp>0
          \ifnum\epsftsize<\epsfrsize\else
             \advance\epsftsize-\epsfrsize \advance\epsfxsize\epsftmp \fi
       \repeat
     \fi
   \else\epsftmp=\epsfrsize \divide\epsftmp\epsftsize
     \epsfysize=\epsfxsize \multiply\epsfysize\epsftmp   
     \multiply\epsftmp\epsftsize \advance\epsfrsize-\epsftmp
     \epsftmp=\epsfxsize
     \loop \advance\epsfrsize\epsfrsize \divide\epsftmp 2
     \ifnum\epsftmp>0
        \ifnum\epsfrsize<\epsftsize\else
           \advance\epsfrsize-\epsftsize \advance\epsfysize\epsftmp \fi
     \repeat     
   \fi
   \ifepsfverbose\message{#1: width=\the\epsfxsize, height=\the\epsfysize}\fi
   \epsftmp=10\epsfxsize \divide\epsftmp\pspoints
   \epsfsetlab%
   \ifincfigs%
     \vbox to\epsfysize{\vfil\hbox to\epsfxsize{%
        \includegraphics{#1}%
        \epsfsetover\hfil}}%
   \else%
     \epsfsetover%
     \vbox to\epsfysize{\hrule\vss\hbox to\epsfxsize{\vrule height
                        \epsfysize\hfil\vrule}\vss\hrule}%
   \fi%
\epsfxsize=0pt\epsfysize=0pt}%

{\catcode`\%=12 \global\let\epsfpercent=
\long\def\epsfaux#1#2:#3\\{\ifx#1\epsfpercent
   \def\testit{#2}\ifx\testit\epsfbblit
      \epsfgrab #3 . . . \\%
      \epsffileokfalse
      \global\epsfbbfoundtrue
   \fi\else\ifx#1\par\else\epsffileokfalse\fi\fi}%
\def\epsfgrab #1 #2 #3 #4 #5\\{%
   \global\def\epsfllx{#1}\ifx\epsfllx\empty
      \epsfgrab #2 #3 #4 #5 .\\\else
   \global\def\epsflly{#2}%
   \global\def\epsfurx{#3}\global\def\epsfury{#4}\fi}%
\def\epsfsize#1#2{\epsfxsize}
\def\ifspace{\ifcat\issp.\else~\fi}
\def\tspace{\futurelet\issp\ifspace}
\def\a{({\it a\kern 1pt})\tspace}
\def\b{({\it b\kern 1pt})\tspace}
\def\c{({\it c\kern 1pt})\tspace}
\def\d{({\it d\kern 1pt})\tspace}
\def\e{({\it e\kern 1pt})\tspace}
\def\f{({\it f\kern 1pt})\tspace}
\def\g{({\it g\kern 1pt})\tspace}
\def\h{({\it h\kern 1pt})\tspace}
\def\i{({\it i\kern 1pt})\tspace}
\def\j{({\it j\kern 1pt})\tspace}
\def\abc#1{({\it #1\kern 1pt})\tspace}

\newcount\ndots
\def\drawline#1#2{\raise 2.5pt\vbox{\hrule width #1pt height #2pt}}

\def\trian{\raise 1.25pt\hbox{$\scriptscriptstyle\triangle$}\nobreak\ }
\def\solidtrian{\raise 1.25pt
\hbox to 3bp{
\hfill}\nobreak\ }
\def\dsolidtrian{\raise 1.25pt
\hbox to 3bp{
\hfill}\nobreak\ }
\def\soliddiamond{\raise 1.25pt
\hbox to 4bp{
\hfill}\nobreak\ }

\def\square{${\vcenter{\hrule height .4pt 
              \hbox{\vrule width .4pt height 3pt \kern 3pt \vrule width .4pt}
          \hrule height .4pt}}$\nobreak\ }

\def\plus{\raise 1.25pt \hbox{$\scriptscriptstyle +$}\nobreak\ }
\def\x{\raise 1.25pt \hbox{$\scriptscriptstyle \times$}\nobreak\ }
\def\legendtable#1{\vbox{\baselineskip=10pt\tabskip=0pt\let\\=\cr\halign{\hfil##\hskip 3pt&##\hfil\cr#1\crcr}}}
\def\lllegend#1 #2 #3{\figlab {#1} {#2} {\legendtable{#3}}}
\def\lrlegend#1 #2 #3{\figlab {#1} {#2} {\llap{\legendtable{#3}}}}
\def\ullegend#1 #2 #3{\figlab {#1} {#2} {\vtop{\hrule height 0pt\legendtable{#3}}}}
\def\urlegend#1 #2 #3{\figlab {#1} {#2} {\llap{\vtop{\hrule height 0pt\legendtable{#3}}}}}

\newdimen\xorigon
\newdimen\yorigon
\newdimen\scaleval
\newdimen\scaleorigon

\def\setxscale#1 #2 #3 #4 #5 {%
    \xorigon=#1\yorigon=#3%
    \scaleval=#2\advance\scaleval by -\xorigon%
    \tempdimen=#5 pt\advance\tempdimen by -#4pt%
    \divide\tempdimen by 1000%
    \divide\scaleval by \tempdimen%
    \scaleorigon=-#4pt\divide\scaleorigon by 1000%
    \multiply\scaleorigon by \scaleval}
\def\xtickup#1 #2{\tempdimen=#1pt\divide\tempdimen by 1000%
    \multiply\tempdimen by \scaleval\advance\tempdimen by \scaleorigon%
    \advance\tempdimen by \xorigon%
    \figlab {\tempdimen} {\yorigon} {\vbox {\hbox to 0pt{\hss #2\hss}%
        \baselineskip=8pt\lineskiplimit=-5pt%
        \hbox to 0pt{\hss \vrule height 3pt\hss}}}}
\def\xtickdown#1 #2{\tempdimen=#1pt\divide\tempdimen by 1000%
    \multiply\tempdimen by \scaleval\advance\tempdimen by \scaleorigon%
    \advance\tempdimen by \xorigon%
    \figlab {\tempdimen} {\yorigon} {\vbox to 0pt {\hbox to 0pt{\hss \vrule height 3pt\hss}%
        \nointerlineskip\vskip 3pt%
        \hbox to 0pt{\hss #2\hss}\vss}}}
\def\nofig#1#2{\leavevmode{\vbox {\hrule \hbox to #1{\vrule height #2 \hfill \vrule} \hrule}} }

    \numberwithin{algorithm}{subsection}
    \floatname{algorithm}{Algoritmo}

\newcommand{\captionfonts}{\small}

\makeatletter  
\long\def\@makecaption#1#2{%
  \vskip\abovecaptionskip
  \sbox\@tempboxa{{\captionfonts #1: #2}}%
  \ifdim \wd\@tempboxa >\hsize
    {\captionfonts #1: #2\par}
  \else
    \hbox to\hsize{\hfil\box\@tempboxa\hfil}%
  \fi
  \vskip\belowcaptionskip}
\makeatother   

\begin{document}
%
\title{Token-DCF: An Opportunistic MAC protocol for Wireless Networks}

\date{February 2, 2012}

\author{\IEEEauthorblockN{Ghazale Hosseinabadi and Nitin Vaidya}
\IEEEauthorblockA{Department of ECE and Coordinated Science Lab.\\
University of Illinois at Urbana-Champaign\\
\{ghossei2,nhv\}@illinois.edu\\\\
Technical Report (February 2, 2012)}}

\maketitle
\IEEEpeerreviewmaketitle

\begin{abstract}
IEEE 802.11 DCF is the MAC protocol currently used in wireless LANs. 802.11 DCF is inefficient due to two types of overhead; channel idle time and collision time. This paper presents the design and performance evaluation of an efficient MAC protocol for wireless networks, called Token-DCF. Token-DCF decreases both idle time and collision time. In Token-DCF, each station keeps track of neighboring links' queue length by overhearing of transmitted packets on the wireless medium. The result is then used to assign privileges to the network stations. A privileged station does not follow the backoff mechanism and transmits immediately after the channel is sensed idle. Our simulation results show that Token-DCF can significantly improve channel utilization, system throughput and channel access delay over 802.11 DCF. 
\end{abstract}
\section{Introduction}
IEEE 802.11 defines the distributed coordination function (DCF) to share the wireless medium among multiple stations. 
DCF employs CSMA/CA with the binary exponential backoff algorithm to resolve channel contention. DCF specifies random backoff, which forces a station to defer its access to the channel for a random period of time. Backoff counter corresponds to the number of idle slots a station has to
wait before its transmission attempt. If multiple stations choose the same backoff, they will attempt to transmit at the same time and collisions will occur. \let\thefootnote\relax\footnotetext{This research is supported in part by National
Science Foundation award CNS 11-17539. Any opinions, findings, and conclusions or recommendations expressed here are those of the authors and do not
necessarily reflect the views of the funding agencies or the U.S. government.}

Two types of overhead are associated with random access protocols. One overhead is channel idle time (e.g. backoff time) which is the time when contending stations are waiting to transmit. Another is collision when multiple stations transmit simultaneously. If there are few contending stations, idle time is the dominant overhead. If there are many contending stations, collision probability increases and becomes the main reason of low channel utilization. In the literature, many MAC protocols have been proposed to reduce the total overhead caused by idle periods and collisions \cite{EnhancedDCF}, \cite{MFS}, \cite{CSMAp*}, \cite{IdleSense}, \cite{pipelining}. 

In this paper, we design an efficient MAC protocol, called \emph{Token-DCF}, in which both idle time and collision time are reduced significantly. In Token-DCF, when a station transmits on the channel, it might give a privilege to one of its neighbors. When a transmission finishes, the privileged station, if there is any, starts transmission after a short period of time, namely SIFS (Short Inter Frame
Space). Non-privileged stations follow the backoff mechanism of 802.11 to access the channel. In this way, the privileged station does not go through the contention resolution phase and grabs the channel immediately. Since in Token-DCF contention resolution is done via assigning tokens, or privileges, idle time and collision time are decreased significantly. 

A scheduling policy is a rule to determine a set of
links to be activated simultaneously at each time instant.
The scheduling policy of 802.11 DCF is CSMA random access, in which a station has to sense the channel as idle for a random period of time before it transmits on the channel. Depending on network configuration, IEEE 802.11 can operate very far from the throughput capacity. Several centralized
and distributed scheduling algorithms were designed
for wireless networks \cite{Tassiulas}, \cite{LQF}, \cite{Walrand}, \cite{NiSrikant} which have better throughput characteristics than 802.11 DCF. Centralized scheduling algorithms rely on a centralized coordinator to manage channel access, which is often not available in distributed networks. Distributed scheduling algorithms do not rely on such a coordinator and each station makes the scheduling decision in a distributed way. 

Token-DCF is fully distributed and does not require any centralized point of coordination. Furthermore, it works for both single hop and multi hop flows. In Token-DCF, a station might schedule its neighbors for transmission on the channel. In this way, each network station performs as a scheduler. Token-DCF is flexible in the sense that it allows different scheduling mechanisms to be used for assigning privileges to network stations. Token-DCF uses an opportunistic approach based on packet overhearing to exchange scheduling information. In Token-DCF, queue length of a station is included in the MAC header of the packets it transmits and is overheard by the neighboring stations. Each station keeps track of queue length of its neighbors. Queue length information is used in the scheduling component of the protocol, which chooses a neighbor of the transmitting station as the privileged station. No extra control packet is transmitted to assign a privilege to a station. Instead, the next privileged station (scheduled station) is specified in the MAC header of data packets being transmitted on the channel. The probability of assigning a privilege is always less than $1$ to allow transmission of newly arrived traffic on the channel as well as imperfections in traffic estimation. This probability is adjusted based on the accuracy of the neighbors' traffic estimation. 

The rest of the paper is organized as follows. We first review some related work is Section \ref{RelatedWork}. We then present our protocol, Token-DCF, in Section \ref{Protocol}. We compare our protocol with IEEE 802.11 in Section \ref{Simulation} and present some conclusions in Section \ref{conclusion}.

\section{Related Work}\label{RelatedWork}
We review four categories of existing work,
\begin{enumerate}
\item Protocols to decrease the idle time and collision time of IEEE 802.11 DCF \cite{EnhancedDCF}, \cite{MFS}, \cite{CSMAp*}, \cite{IdleSense}, \cite{pipelining}, \cite{CHAIN}.
\item Token passing MAC protocols \cite{tokenBus}, \cite{TokenRing}, \cite{robustToken}, \cite{wirelessToken}.
\item Centralized scheduling algorithms for wireless networks \cite{Tassiulas}, \cite{LQF}.
\item Distributed throughput optimal CSMA protocols \cite{Walrand}, \cite{NiSrikant}, \cite{THKim}.
\end{enumerate}

\subsection{Enhancing 802.11 DCF}   
\cite{EnhancedDCF} modifies the backoff algorithm of the IEEE 802.11 MAC protocol and derives the contention window size that maximizes network throughput. The backoff window size is tuned at run time to increase the throughput. In this protocol, in light and medium load conditions, in which the window size defined in 802.11 DCF is sufficient to guarantee low collision probabilities, the standard
backoff algorithm is generally adopted. On the other hand,
when the network congestion increases, a contention window with the right size for that
load condition is used.

Model-based frame scheduling (\emph{MFS}) is presented in \cite{MFS}. In MFS, each
station estimates the current network status by keeping track
of the number of collisions it encounters between its two consecutive
successful frame transmissions, and, based on the
the estimated information, computes the current network
utilization. The result is then used to determine a scheduling
delay that is introduced, with the objective of avoiding
collision, before a station attempts for transmission of its
pending frame.

\cite{CSMAp*} considers a network in which all stations become simultaneously backlogged at some point in time and designs \emph{CSMA/p$^*$} to find the optimal backoff distribution according to which every station chooses a contention slot. 
In \emph{Idle Sense} \cite{IdleSense}, each host observes the mean
number of idle slots between transmission attempts to dynamically
control its contention window. Idle Sense enables each host to estimate its frame
error rate, which is used for switching to the right bit
rate. \emph{Implicit pipelining} \cite{pipelining} parallelizes part of the contention resolution time and packet transmission time. It partially hides channel idle overhead and reduces collision overhead.

\cite{CHAIN} presents \emph{CHAIN}, in which clients maintain a precedence relation among one
another, and a client can immediately transmit a new packet
after it overhears a successful transmission of its predecessor,
without going through the regular contending process. When
the network load is low, CHAIN behaves similar to DCF; But
when the network becomes congested, clients automatically start
chains of transmissions to improve efficiency. CHAIN requires transmission of control packets between an access point and its stations periodically, which adds overhead to the protocol. Furthermore, during each scheduling period, the specified precedence relation is fixed and does not adapt to traffic changes during that period. 

\subsection{Token passing MAC protocols}
Token passing is a medium access method where a short packet called a \emph{token} is passed between stations that authorizes the station to transmit. In token passing protocols, stations take turns in transmitting by passing the token from station to station. Stations that have data frames to transmit must first acquire the token before they can transmit them. A station can only send data if it possesses the token, thus avoiding collisions. 
Token passing schemes provide round-robin scheduling method. The advantage over contention based medium access is that collisions are eliminated, and that the available bandwidth can be fully utilized without idle time when demand is heavy. The disadvantage is that even when demand is light, a station wishing to transmit must wait for the token, increasing latency.

The IEEE 802.4 Token Bus protocol \cite{tokenBus} is a well-known example of token passing protocols. Token bus protocol is based on a broadcast
medium (broadband coaxial cable), which connects all nodes
to each other. The token is passed among a logical ring of
stations attached to the cable. The stations sort themselves
for order of token passing by their MAC addresses.

IEEE has standardized another token passing MAC protocol for wired networks, called token ring (IEEE 802.5) \cite{TokenRing}. Stations on a token ring LAN are logically organized in a ring topology with data being transmitted sequentially from one ring station to the next with a control token circulating around the ring controlling access. In token ring standard, token is passed around a ring and whichever station holds the token is allowed to transmit before putting the token back on the ring.

The Wireless Token Ring Protocol (WTRP) \cite{wirelessToken} is a
token bus protocol, derived from IEEE 802.4. WTRP presents a token passing MAC protocol for wireless networks. When token passing is
to be used in a WLAN, the characteristics of the wireless medium, such as connectivity loss, network partitioning and token loss, raise additional token management issues. WTRP is capable of recovering from token loss
and duplication, and dealing with changes in network connectivity and membership. The principal
modifications of 802.4 that are introduced by WTRP address
the partial connectivity issues that arise in wireless networks.

\cite{robustToken} designs another token passing MAC protocol for wireless networks, called high frequency token protocol (HFTP). HFTP is based on WTRP, but adds two new mechanisms: token relaying and a ring merging procedure. Token relaying deals with the situation when a station attempts to pass the token to its successor, but fails to receive acknowledgement due to a link outage. HFTP will
attempt to find an indirect path to its successor rather than reconnecting
the ring to exclude that node. This requires new
mechanisms to find and to use token relay nodes. HFTP also differs from WTRP in its mechanism for merging
rings that come into range of each other. This can occur after
a network that was partitioned regains connectivity.

\subsection{Centralized Scheduling Algorithms}\label{centerSched}
The first throughput optimal scheduling algorithm was introduced
in the seminal work of Tassiulas and Ephremides \cite{Tassiulas}.
The proposed algorithm is a centralized algorithm known as
Backpressure. In Backpressure algorithm the schedule at each time slot $t$ is determined by
\begin{equation}\label{BPgeneral}
\vec{r}(t)=\mbox{argmax}_{\vec{r} \in { \mathcal{R}}}  \bigg{[}\sum_{(i,j)} (q_i-q_j)r_{ij}\bigg{]}
\end{equation}
For each link $(i,j)$ from station $i$ to station $j$, $(q_i-q_j)$ denotes its queue differential and $r_{ij}$ denotes its rate. $\mathcal{R}$ is the convex hull of the capacity region. In Backpressure, at
each time slot, the set of non-conflicting links that maximizes
the above sum is activated. Backpressure is a centralized throughput optimal
scheme, which is capable of scheduling all feasible
flow arrivals while maintaining the network stable. When flows are single hop, i.e., communication is between adjacent stations, then
the backpressure algorithm reduces to 
\begin{equation}\label{BP}
\vec{r}(t)=\mbox{argmax}_{\vec{r} \in { \mathcal{R}}}  \bigg{[}\sum_{(i,j)} q_i(t)r_{ij}\bigg{]}
\end{equation}
Another important scheduling policy which has been observed to achieve $100\%$ throughput in most practical wireless networks is longest-queue-first scheduling, also called greedy maximal scheduling \cite{LQF}. LQF makes scheduling decisions based on queue length information as follows. It starts with an empty schedule. It first adds the link with the largest queue length to the schedule. It then looks for the link with the largest queue length among the remaining links. This chosen link will be added to the schedule if this addition creates a feasible schedule, i.e. the set of added links satisfies the SINR constraints, or it is discarded otherwise. This process continues until no link is left. 

A centralized scheduler requires a central authority
to determine the schedule. In a distributed wireless
network such a central authority does not necessarily exist.
Consequently, various distributed scheduling schemes were
designed for wireless networks that might not be throughput
optimal but are simpler than a centralized algorithm and can be
implemented in a large scale wireless network. 
\subsection{Throughput optimal CSMA}
Recently, it has been shown that distributed carrier sense
multiple access (CSMA) algorithms can achieve throughput optimality under certain network models and assumptions \cite{Walrand}, \cite{NiSrikant}, \cite{THKim}. Jiang and Walrand \cite{Walrand} proposed a distributed adaptive random access CSMA algorithm, in which the back-off time of a link is an exponentially distributed random variable. The mean of this random variable changes over time and its dynamic is determined by the queue length of the link. Their algorithm achieves throughput-optimality under
the assumption of continuous-time backoff
duration (zero probability of collision) and continuous-time transmission duration.

Ni and Srikant \cite{NiSrikant} designed a distributed CSMA/CA protocol for achieving maximum throughput in a discrete-time setting. In their work, the model of an idealized CSMA protocol with continuous backoff times, under which collisions cannot occur, is relaxed. Collisions of data packets are avoided through the exchange of
control messages, where control messages might collide. The optimality in protocols designed in \cite{Walrand} and \cite{NiSrikant} is established under the ideal carrier
sensing assumption, i.e., each link can precisely sense the presence
of other active links in its neighborhood. \cite{THKim} investigates the achievable throughput of the CSMA algorithm
under imperfect carrier sensing. Their main result is that CSMA
can achieve an arbitrary fraction of the capacity region if certain
access probabilities are set appropriately.

Channel access method in throughput-optimal CSMA protocols is random access in which contention among the stations for accessing the channel is resolved through the backoff mechanism. This results in non-trivial backoff overhead (idle time) in these protocols.
\section{Token-DCF design}\label{Protocol}
In this section, we first provide a high-level overview of Token-DCF and then detail the scheduler signaling and scheduling algorithm. 
\subsection{Overview}
At a high level, the operation of Token-DCF is described
as follows. Token-DCF runs a distributed scheduling protocol, where a privilege might be assigned by a transmitting station to one of its neighbors. In each transmission, the transmitting station
might select one of its neighbors to have a higher priority for the next transmission. Selection mechanism is based on flow queue lengths. When a transmission finishes, the station with a privilege, called \textit{\texttt{privileged}}, starts transmission after a short period of time, SIFS (Short Inter Frame
Space), if the channel is sensed idle. 

Token-DCF is implemented in the MAC layer of the protocol stack. Scheduling information is embedded in the MAC header of the data packets and is transferred to the stations via overhearing. Token-DCF reduces signaling overhead in its scheduling component compared to central scheduling algorithms. Each station maintains queue length of the neighboring stations. The queue lengths are used in the scheduling component to select the privileged station for the next transmission. Transmitting station announces the privileged station in the \textit{\texttt{privileged}} field of the MAC header of the data packets it transmits. By overhearing of these packets, the privileged station is informed that it has a higher priority for the next transmission. When a transmission finishes, the privileged station can start transmission after SIFS, if the channel is sensed idle. Note that in multi-hop networks, at each time instance, several privileged stations might be present in the network, since in multi-hop networks, non-interfering transmitters transmit at the same time and each of them assigns a privilege to one of their neighbors.

Signaling mechanism in the scheduling component of Token-DCF is done via embedding the scheduling information in the header of data packets by the source station and overhearing of the packets to retrieve such information by the neighboring stations. When a packet is transmitted, the station that will have higher priority for the next transmission, the \textit{\texttt{privileged}} station, and the queue length of the transmitter are embedded in the MAC header of the packet. Once a packet is received or overheard, queue length of the source of the packet is saved by the receiving or overhearing station. Furthermore, a station that receives or overhears a packet, checks the \textit{\texttt{privileged}} field of the MAC header of the packet to find if it is chosen to be the next privileged station. In Token-DCF, no central scheduler is deployed in the network and no extra control messages are transmitted to find and disseminate a schedule. Collecting the information needed for scheduling, assigning a privilege to one of the neighbors and obtaining the privilege by the \textit{\texttt{privileged}} station are all done via receiving or overhearing of data packets. 
\begin{figure}[t]
\includegraphics [width= 90 mm]{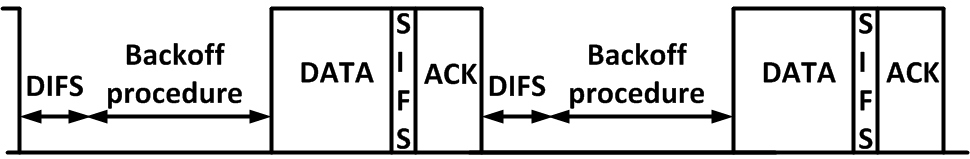}\label{DCF}
\caption{Access method of IEEE 802.11 DCF}\label{DCF}
\end{figure}

Token-DCF has two major components: (1) A method to reduce the idle time of the backoff mechanism. (2) A scheduling algorithm to determine which neighbor is chosen as the privileged station.
\subsection{Reducing idle time}
Token-DCF reduces the idle time of the backoff mechanism by assigning privileges to network stations. When a station transmits data packets, it might give higher priority for the next transmission to one of its neighbors. A transmitting station gives a high priority to one of its neighbors with probability \textit{\texttt{p}}. With probability $1-$\textit{\texttt{p}}, no station is given a higher priority. As we will explain in Section \ref{SchedAlgo}, the scheduling algorithm of Token-DCF determines which neighbor is chosen as the privileged station, i.e., the station with a higher priority. When a transmission finishes, a station that has a privilege starts transmission after short period of time, 
SIFS, if the channel is sensed idle. Non-privileged stations follow the backoff mechanism of IEEE 802.11 to access the wireless medium. Backoff mechanism of 802.11 DCF is shown in Figure \ref{DCF}. In this mechanism, after a transmission finishes, the station senses the channel after DIFS interval and if the channel is sensed idle, it waits for a random contention time: it chooses backoff
$b$, an integer distributed uniformly in the window $[0, CW]$
and waits for $b$ time slots before attempting to transmit. 

Channel access mechanism of our protocol, Token-DCF, is shown in Figure \ref{TokenDCF}. In Token-DCF, when the channel becomes idle, the privileged station, if there is any, starts transmission on the channel immediately, and non-privileged stations have to defer backoff count down till when transmission of the privileged station finishes. The process of giving a privilege by a transmitting station to one of its neighbors repeats in each transmission. Whenever a privileged station transmits on the channel, the idle time of the channel is only SIFS. On the other hand, in IEEE 802.11 protocol, the channel idle time between two consecutive transmissions is equal to DIFS plus random backoff duration.    
\subsection{Scheduling algorithm}\label{SchedAlgo}
The scheduling algorithm of Token-DCF provides a mechanism for choosing the privileged stations. Information needed in the scheduling component of the protocol is embedded in the MAC header of data packets. Such information includes queue length of transmitter of the packet and the next privileged station. Each station keeps track of neighbor's queue length in order to enable neighbor scheduling. In Token-DCF, when a station transmits, it acts as a scheduler as well and with probability \textit{\texttt{p}} gives a higher priority for the next transmission to one of its neighbors. This technique removes the need for a separate scheduler as well as transmission of control messages between the scheduler and network stations. In central scheduling algorithms, scheduler component and network hosts must exchange control information to coordinate the schedule. As a trade off, our approach is opportunistic and uses message overhearing to exchange the information needed in the scheduling component. 
\begin{figure}[t]
\includegraphics [width= 90 mm]{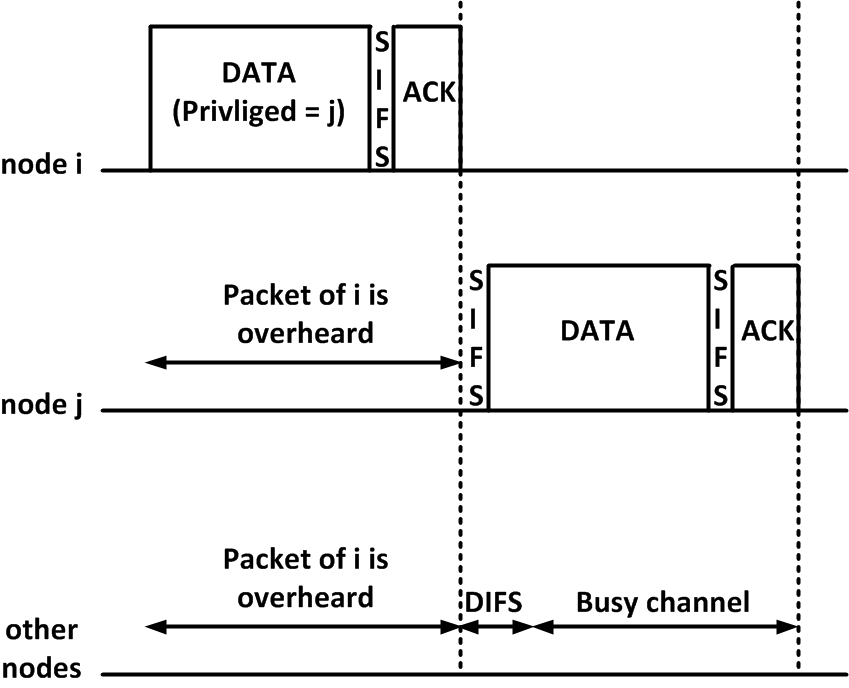}\label{TokenDCF}
\caption{Access method of Token-DCF protocol}\label{TokenDCF}
\end{figure}

Different mechanisms can be used to choose the privileged station. In this paper, we presently consider only single hop
flows (i.e., sender and receiver are adjacent nodes), but our ideas can be extended to multi-hop flows as well.
A station might choose the neighbor with the largest $q_ic_{ij}$ as the next privileged station, where $q_i$ is queue length of transmitter of link $(i,j)$ and $c_{ij}$ is the capacity of link $(i,j)$. If this policy is implemented as the scheduling component of the protocol, a transmitting node should announce its queue length, $q_i$, as well as its link capacity, $c_{ij}$, in the packets it transmits. In single hop networks, if every station overhears packets of every other station, this policy implements backpressure scheduler, explained in Section \ref{centerSched}. In single hop networks, data transmission of any two links interferes with each other and as a result, at each time instance, at most one link can be scheduled for transmission. If network is single hop and every station overhears every other transmission, each station knows queue length $q_i$ and capacity $c_{ij}$ of other network links and schedules the link with the largest $q_ic_{ij}$ as the privileged link. In single hop networks, the link with largest $q_ic_{ij}$ is the one that maximizes Equation (\ref{BP}). In practice, $c_{ij}$ may be approximated by the transmission rate used by the MAC-layer rate control algorithm.   

Another scheduling policy is to pick the link with the longest queue. In single hop networks, when every station overhears every transmission, this policy implements longest-queue-first (LQF) as the scheduling component of Token-DCF. LQF is throughput optimal if the so called local pooling condition is satisfied \cite{LQF}. In our simulations, we have used LQF as the scheduling component of Token-DCF, and all transmissions occur at a fixed rate. 

\subsection{Protocol details}
Procedure \ref{proc:init} sets the initial value of protocol parameters. 
A station that is going to transmit on the channel, with probability \textit{\texttt{p}}, chooses one of its neighbors to have a higher priority for transmission. With probability $1-$\textit{\texttt{p}}, no station is chosen to have a privilege. 
\textit{\texttt{p}} is initially set to zero and changes during the protocol execution in order to adapt the probability of giving a privilege to neighbors. \textit{\texttt{active}} denotes the set of neighbors of a station that has transmitted on the channel during the current scheduling period and the transmission is overheard by the station. The station itself, \textit{\texttt{myId}}, is also included in the set \textit{\texttt{active}}. When a station transmits, it might give a privilege to one of the stations in the set \textit{\texttt{active}}. By including \textit{\texttt{myId}} in the set \textit{\texttt{active}}, a station might choose itself as the privileged station. Each station keeps track of the transmissions on the channel by overhearing of the packets. \textit{\texttt{success}} denotes the number of transmissions from the set \textit{\texttt{active}}. \textit{\texttt{fail}} denotes the number of transmissions in which the sender of the packet is not in the set \textit{\texttt{active}}. Protocol parameters are reset to initial values each \textit{\texttt{period}} seconds. Protocol parameters are reset periodically in order to prevent stale information making the protocol unfair. An alternative to this method (i.e., resetting the initial values) is to use moving average for adapting parameter values during the protocol execution.   

\algsetup{indent=2em}
\floatname{algorithm}{}
\begin{algorithm}[h!]
\caption{Initialization at station \textit{\texttt{myId}}}\label{proc:init}
\begin{algorithmic}[1]
\STATE \textit{\texttt{p}} $=$ $0$
\STATE \textit{\texttt{active}} $=\{ \textit{\texttt{myId}}\}$  
\STATE \textit{\texttt{success}} $=$ $0$
\STATE \textit{\texttt{fail}} $=$ $0$
\STATE call Initialization after \textit{\texttt{period}}
\end{algorithmic}
\end{algorithm}

Procedure \ref{proc:transmitPkt} is executed right before a packet is transmitted on the channel. If the packet is a MAC data packet, the station might give a privilege to one of its neighbors. The mechanism of assigning a privilege or transmitting as the privileged station is not used when control packets are transmitted. In this way, the transmission of non-data packets such as ARP packets or routing packets are not affected by our protocol. The station chooses a \textit{\texttt{privileged}} station with probability \textit{\texttt{p}}, where \textit{\texttt{privileged}} is the station in the set \textit{\texttt{active}} with the longest queue. With probability $1-$\textit{\texttt{p}}, no station is given a privilege. If a station chooses itself as the \textit{\texttt{privileged}}, it sets its \textit{\texttt{flag}} to $1$. Otherwise, \textit{\texttt{flag}} is set to $0$. \textit{\texttt{flag}} equals to $1$ means that the station has a privilege for the next transmission on the channel. Procedure \ref{proc:update}, called Adapt, is then called to update \textit{\texttt{success}}, \textit{\texttt{fail}} and \textit{\texttt{p}}. 
\algsetup{indent=2em}
\floatname{algorithm}{}
\begin{algorithm}[h!]
\caption{Transmit a packet}\label{proc:transmitPkt}
\begin{algorithmic}[1]
\IF {it is a MAC data packet}
\STATE with probability \textit{\texttt{p}}
\STATE \textit{\texttt{privileged}} = station with the longest queue in \textit{\texttt{active}}
\IF {\textit{\texttt{privileged}} $==$ \textit{\texttt{myId}}}
\STATE \textit{\texttt{flag}} $=$ $1$
\ELSE
\STATE \textit{\texttt{flag}} $=$ $0$
\ENDIF
\STATE Adapt
\ELSE
\STATE \textit{\texttt{privileged}} $=$ \textit{\texttt{null}} 
\ENDIF
\end{algorithmic}
\end{algorithm}

Procedure \ref{proc:OH} is called when a packet is received or overheard. Since the wireless channel is a shared medium, station $i$ might overhear packets that are not intended for it, i.e., packets with destination address different from $i$. If the station is chosen to be the \textit{\texttt{privileged}} in the received or overheard packet, it sets its \textit{\texttt{flag}} to $1$. Otherwise, \textit{\texttt{flag}} is set to $0$. The station then calls Adapt, Procedure \ref{proc:update}, in which \textit{\texttt{success}}, \textit{\texttt{fail}} and \textit{\texttt{p}} are updated. The station also saves the queue length of \textit{\texttt{src}} in its \textit{\texttt{qLen}}.   
\algsetup{indent=2em}
\floatname{algorithm}{}
\begin{algorithm}[h!]
\caption{Receive or Overhear a packet from station \textit{\texttt{src}}}\label{proc:OH}
\begin{algorithmic}[1]
\IF {\textit{\texttt{privileged}} $==$ \textit{\texttt{myId}}}
\STATE \textit{\texttt{flag}} $=$ $1$
\ELSE
\STATE \textit{\texttt{flag}} $=$ $0$ 
\ENDIF
\STATE Adapt
\STATE \textit{\texttt{qLen[src]}} $=$ queue length of \textit{\texttt{src}}
\end{algorithmic}
\end{algorithm}

Procedure \ref{proc:startBackoffTimer} is executed when a station starts or resumes its backoff timer. If the station has higher priority, i.e., \textit{\texttt{flag}} $==$ $1$, and the packet is a MAC data packet, the backoff duration is set to SIFS. Otherwise, the backoff duration is chosen to be DIFS plus random number of time slots, similar to 802.11 DCF. There are alternatives to this approach, for example the privileged station might choose a smaller backoff compared to non-privileged stations. This approach will decrease the probability of collision when there are multiple privileged stations. Recall that in multihop networks, at each time instant more than one privileged node might exist in the network.   

\algsetup{indent=2em}
\floatname{algorithm}{}
\begin{algorithm}[h!]
\caption{Start or resume backoff timer}\label{proc:startBackoffTimer}
\begin{algorithmic}[1]
\IF {\textit{\texttt{flag}} $==$ $1$ \&\& packet is a MAC data packet}
\STATE schedule backoff timer for SIFS
\ELSE
\STATE schedule backoff timer for DIFS $+$ random number of time slots 
\ENDIF
\end{algorithmic}
\end{algorithm}

When the backoff timer expires, \textit{\texttt{flag}} is reset to zero. In this way, a privileged station has the privilege to transmit only one packet immediately after the last transmission finishes. In case the packet is lost, the station does not have the privilege for retransmission of the packet and will follow the backoff mechanism to access the channel. When a host detects a failed transmission (it does
not receive the ACK of a frame), it executes the exponential backoff algorithm---it doubles contention window $CW$ ($CW$ may vary between $CW_{min}$ and $CW_{max}$).  

As explained before, when a packet is transmitted or received, Adapt (Procedure \ref{proc:update}) might be called, in order to update the value of \textit{\texttt{success}}, \textit{\texttt{fail}} and \textit{\texttt{p}}. Station $i$ calls Adapt when it transmits a packet or when it receives or overhears a transmission. If transmitter of the packet, \textit{\texttt{src}}, does not belong to the set \textit{\texttt{active}}, \textit{\texttt{fail}} is increased by one and \textit{\texttt{src}} is added to the set \textit{\texttt{active}}. In this case, the station that receives or overhears the packet, has not received any transmission from \textit{\texttt{src}} during the current scheduling period. Otherwise, If \textit{\texttt{src}} belongs to the set \textit{\texttt{active}}, \textit{\texttt{success}} is increased by $1$. Recall that the set \textit{\texttt{active}} is reset every \textit{\texttt{period}} seconds.

Ratio of \textit{\texttt{success}} to \textit{\texttt{success}}+\textit{\texttt{fail}} is then considered to adapt $p$. If \textit{\texttt{ratio}} is larger than a threshold, \textit{\texttt{maxRatio}}, and enough number of transmissions have happened (i.e., \textit{\texttt{success}}$+$\textit{\texttt{fail}} $>=$ \textit{\texttt{maxNum}}) \textit{\texttt{p}} is increased by $\delta$ and \textit{\texttt{success}} and \textit{\texttt{fail}} are reset to $0$. We note that \textit{\texttt{p}} is increased up to a threshold, \textit{\texttt{maxP}}. It is reasonable to choose \textit{\texttt{maxP}} less than $1$ in order to always give a chance to stations not in \textit{\texttt{active}} to be able to transmit on the channel. If \textit{\texttt{ratio}} is less than a threshold, \textit{\texttt{minRatio}}, while enough number of transmissions have happened (i.e., \textit{\texttt{success}}$+$\textit{\texttt{fail}} $>=$ \textit{\texttt{maxNum}}), \textit{\texttt{p}} is decreased by $\delta$ and \textit{\texttt{success}} and \textit{\texttt{fail}} are reset to $0$.  
\algsetup{indent=2em}
\floatname{algorithm}{}
\begin{algorithm}[h!]
\caption{Adapt}\label{proc:update}
\begin{algorithmic}[1]
\IF {\textit{\texttt{src}} $\notin$ \textit{\texttt{active}}}
\STATE \textit{\texttt{fail}} $++$
\STATE add \textit{\texttt{src}} to \textit{\texttt{active}} 
\ELSE
\STATE \textit{\texttt{success}} $++$
\ENDIF
\IF {(\textit{\texttt{success}}$+$\textit{\texttt{fail}} $>=$ \textit{\texttt{maxNum}})}
\STATE \textit{\texttt{ratio}} $=$ \textit{\texttt{success}} $/$ $($\textit{\texttt{success}}$+$\textit{\texttt{fail}}$)$
\IF {(\textit{\texttt{ratio}} $>=$ \textit{\texttt{maxRatio}})}
\IF {(\textit{\texttt{p}} $<=$ \textit{\texttt{maxP}})}
\STATE \textit{\texttt{p}} $=$ \textit{\texttt{p}} $+$ $\delta$
\ENDIF
\STATE \textit{\texttt{success}} $=$ $0$
\STATE \textit{\texttt{fail}} $=$ $0$
\ENDIF
\IF {(\textit{\texttt{ratio}} $<=$ \textit{\texttt{minRatio}})}
\IF {(\textit{\texttt{p}} $>=$ $\delta$)}
\STATE \textit{\texttt{p}} $=$ \textit{\texttt{p}} $-$ $\delta$
\ENDIF
\STATE \textit{\texttt{success}} $=$ $0$
\STATE \textit{\texttt{fail}} $=$ $0$
\ENDIF
\ENDIF
\end{algorithmic}
\end{algorithm}
There are other alternatives for adapting protocol parameters. For example, different moving average techniques (e.g., weighted, exponential, $\cdots$) can be used to adapt the protocol parameters.  
\section{Evaluation}\label{Simulation}
We simulate Token-DCF and 802.11g \cite{Amenment2007} in ns-2 to measure and compare performance of these two MAC protocols. The network is a wireless ad hoc network in which transmitting stations are placed uniformly at random in a square area. Flows might be single hop or multi hop. In the case of single hop flows, the receiver of each flow is placed at a distance of $100m$ from the transmitter of the flow. This means that if transmitter is placed at point $(x,y)$, receiver is placed at point $((x+100) \mbox{ mod } d,y)$, where $d$ is area length. In case of multi-hop flows, receiving stations are placed uniformly at random in the area. We run the simulations for different network sizes, including single-hop and multi-hop networks. The effective transmission range in the simulations is limited to $250$ meters and carrier sense range is limited to $550$ meters. IEEE 802.11 RTS/CTS mechanism is turned off. Two-ray ground radio propagation model is assumed. Each simulation lasts for $30$ seconds and the presented results are averaged over $5$ runs. In each run, a different random network topology is considered. We measure the results of Token-DCF and 802.11 DCF in terms of aggregate throughput, average access delay, channel idle time and collision frequency. Table \ref{values} reports the configuration parameter values of the wireless network analyzed in this section. Table \ref{protocolVals} reports the parameter values of Token-DCF chosen in the simulations.  
\begin{table}[h] 
\begin{center}
  \begin{tabular}{| l | c | }
    \hline
    SIFS & 10 $\mu$sec\\ \hline
    DIFS & 28 $\mu$sec\\ \hline
    slot time & 9 $\mu$sec\\ \hline
    phy preamble & 16 $\mu$sec\\ \hline
    bit rate & 54 Mbps\\ \hline
    CWmin & 16\\ \hline
    CWmax & 1024\\ 
    \hline
  \end{tabular}
  \caption{WLAN configuration}
  \label{values}
  \end{center}
\end{table}
\begin{figure}[t]
\begin{center}
\includegraphics [width= 96 mm]{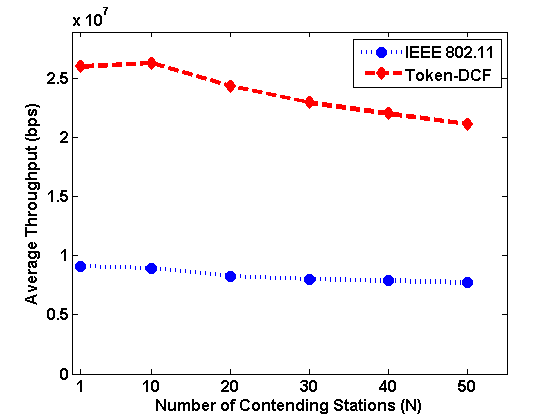}
\end{center}
\caption{System throughput (area=150mx150m, packet size=500B)}\label{thr_A150_ps500}
\end{figure}
\begin{table}[t] 
\begin{center}
  \begin{tabular}{| l | c | }
    \hline
    \textit{\texttt{minRatio}} & 0.2 \\ \hline
    \textit{\texttt{maxRatio}} & 0.8\\ \hline
     \textit{\texttt{maxNum}} & 20 \\ \hline
    $\delta$ & 0.1 \\ \hline
    \textit{\texttt{maxP}} & 0.9 \\ \hline
    \textit{\texttt{period}} & 0.1 sec\\ 
    \hline
  \end{tabular}
  \caption{Token-DCF parameters}
  \label{protocolVals}
  \end{center}
\end{table}
\subsection{Performance evaluation in single-hop networks }
\begin{figure}[t]
\begin{center}
\includegraphics [width= 96 mm]{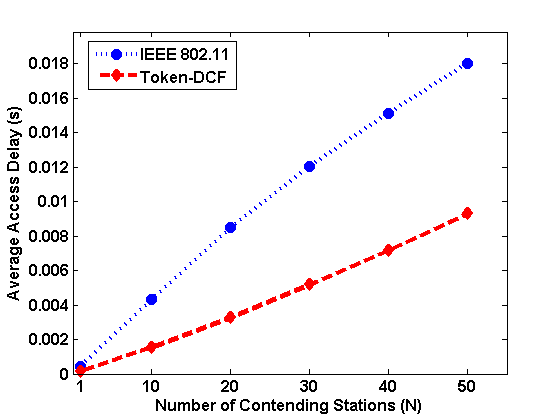}
\end{center}
\caption{Average access delay (area=150mx150m, packet size=500B)}\label{delay_A150_ps500}
\end{figure}

\begin{figure}[t]
\begin{center}
\includegraphics [width= 96 mm]{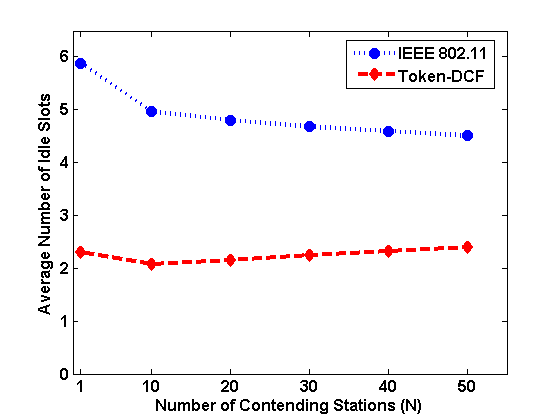}\label{numIdleSlots_A150_ps500}
\end{center}
\caption{The average number of idle slots before each media access (area=150mx150m, packet size=500B)}\label{numIdleSlot_A150_ps500}
\end{figure}

\begin{figure}[t]
\begin{center}
\includegraphics [width= 96 mm]{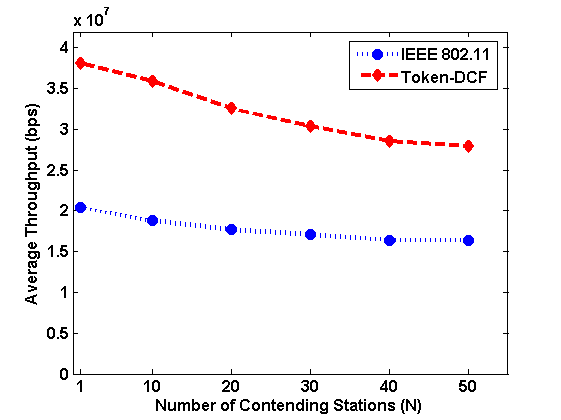}
\end{center}
\caption{System throughput (area=150mx150m, packet size=1500B)}\label{thr_A150_ps1500}
\end{figure}

\begin{figure}[t]
\begin{center}
\includegraphics [width= 96 mm]{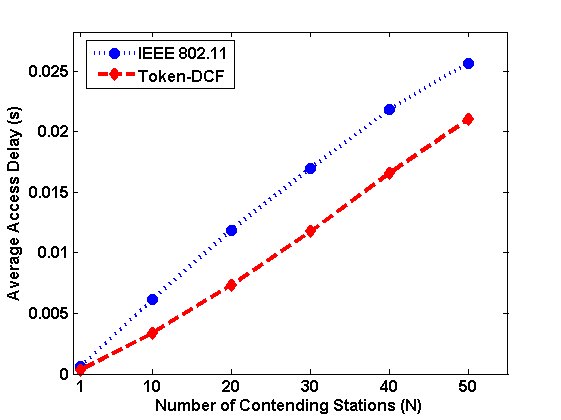}
\end{center}
\caption{Average access delay (area=150mx150m, packet size=1500B)}\label{delay_A150_ps1500}
\end{figure}

Figures \ref{thr_A150_ps500}-\ref{delay_A150_ps1500} plot the performance parameters in a single-hop network. The size of the network is $150m$x$150m$. Traffic is full buffer CBR, meaning that there is always backlogged traffic in the transmission queue of each link. Transmission queue of each link holds up to $50$ packets and when the buffer is full, newly arrived packets get dropped. The payload size is 500 bytes and all flows are single-hop.

The aggregate throughput of 802.11 DCF and Token-DCF is presented in Figure \ref{thr_A150_ps500}. As we can see, throughput gain obtained by Token-DCF compared to IEEE 802.11 in Figure \ref{thr_A150_ps500} is a factor of $2.7-2.9$. Figure \ref{delay_A150_ps500} shows the average access delay of the two protocols. Access delay is defined as the delay between the time a packet arrives at the MAC layer and the time the source of the packet receives acknowledgment from the destination. Access delay of a packet consists of the waiting time before transmitting on the channel and the time spent in packet retransmissions. As we can see in Figure \ref{delay_A150_ps500}, access delay is smaller in Token-DCF by a factor of $0.35-0.51$. Token-DCF has a much shorter idle time compared to IEEE 802.11 DCF. Furthermore, many retransmissions are avoided because of reduced collision frequency. 
Figure \ref{numIdleSlot_A150_ps500} presents the average number of idle slots before each media access. Token-DCF has shorter channel idle time, because in Token-DCF, a privileged station accesses the channel immediately after the latest transmission finishes. In this way, channel stays idle only for SIFS seconds, instead of DIFS plus random backoff duration. We note that the average number of idle slots in Token-DCF is not zero. The reason is that with a non-zero probability, no station is chosen as the privileged station for the next transmission. In such a case, stations follow the backoff mechanism of 802.11 DCF to get an access for transmission on the medium. 

Collision frequency of 802.11 DCF and Token-DCF is shown in Figure \ref{colFreq_A150_ps500}. Collision frequency is defined as the number of times a transmission fails due to collision normalized by the total number of transmissions (counting retransmissions as well). Figure \ref{colFreq_A150_ps500} indicates that Token-DCF has much lower collision frequency than 802.11 DCF. In a single-hop network, at each time instant, at most one station successfully transmits on the media and as a result, there is at most one privileged station at each time instant. Recall that when a station transmits, it might choose one of its neighbors as the privileged station. Since a privileged station does not follow the backoff mechanism of 802.11 DCF, the transmission by a privileged station does not collide with any other transmission in a single-hop network. This reduces the collision frequency of the protocol. Reducing the idle time and collision time of the channel increases throughput and decreases media access delay. As we can see in Figure \ref{colFreq_A150_ps500}, with greater number of contending stations, the collision frequency in both Token-DCF and 802.11 DCF increases. Token-DCF has non-zero collision frequency, because with probability $1-p$, stations implement backoff mechanism for contention resolution, which might cause collisions.

Figures \ref{thr_A150_ps1500} and \ref{delay_A150_ps1500} show the throughput and access delay versus number of transmitters in a network of size $150m$x$150m$ where the packet size is 1500 bytes. Traffic type is full buffer CBR. As we can see in these figures, throughput gain is about $1.7-1.9$ and access delay is reduced by a factor of $0.53-0.81$. When packet size is $1500$ bytes, the throughput gain is lower since
the efficiency of DCF increases with packet size. The overhead per successful packet transmission, $T_{oh}$, is equal to the sum of the channel idle time and collision time. We denote the packet transmission time by $T_{tr}$. $T_{tr}$ is equal to the sum of DIFS, transmission time of the data packet, SIFS and transmission time of the acknowledgement. Then, the efficiency of DCF can be defined as
\begin{equation}
efficiency = \frac{T_{tr}}{T_{oh}+T_{tr}}
\end{equation}
$T_{oh}$ is not a function of the packet size and does not change when packet size changes. On the other hand, $T_{tr}$ increases when packet size increases. This results in larger \emph{efficiency} when packet size is 1500 bytes.
\begin{figure}[t]
\begin{center}
\includegraphics [width= 96 mm]{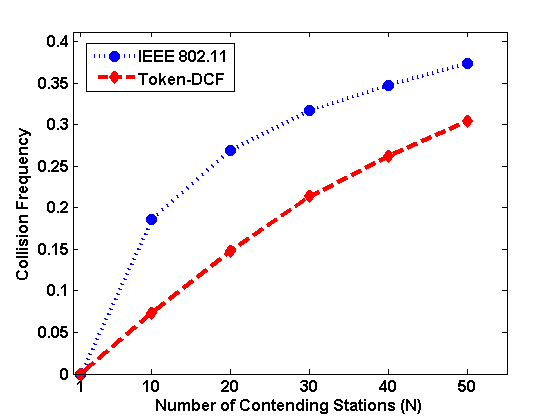}\label{colFreq_A150_ps500}
\end{center}
\caption{Collision frequency (area=150mx150m, packet size=500B)}\label{colFreq_A150_ps500}
\end{figure}   

\subsection{Performance evaluation in multihop wireless networks}
\begin{figure}[t]
\begin{center}
\includegraphics [width= 96 mm]{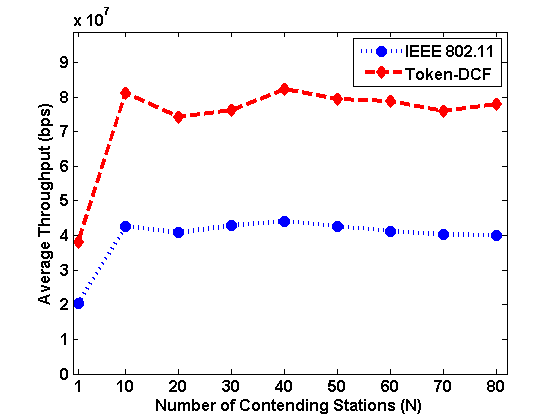}
\end{center}
\caption{System throughput (area=800mx800m)}\label{thr_A800_ps1500}
\end{figure}

\begin{figure}[t]
\begin{center}
\includegraphics [width= 96 mm]{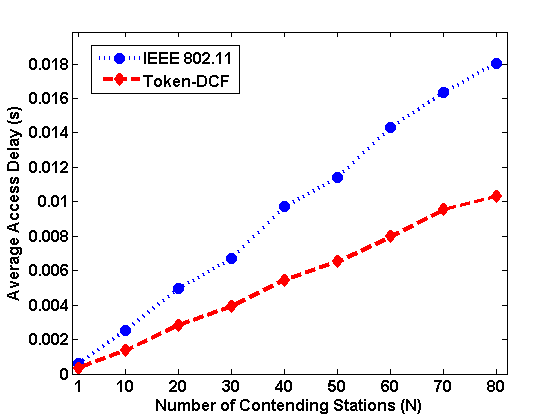}
\end{center}
\caption{Average access delay (area=800mx800m)}\label{delay_A800_ps1500}
\end{figure}

\begin{figure}[t]
\begin{center}
\includegraphics [width= 96 mm]{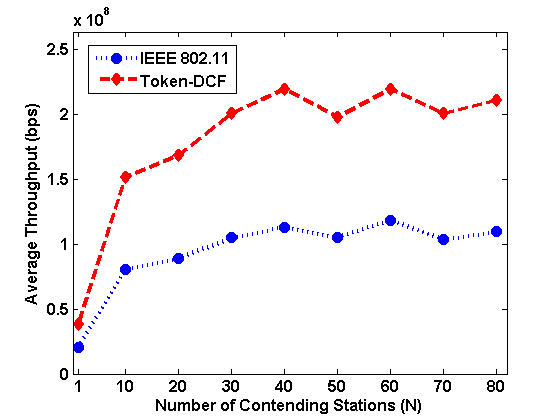}
\end{center}
\caption{System throughput (area=1500mx1500m)}\label{thr_A1500_ps1500}
\end{figure}

\begin{figure}[t]
\begin{center}
\includegraphics [width= 96 mm]{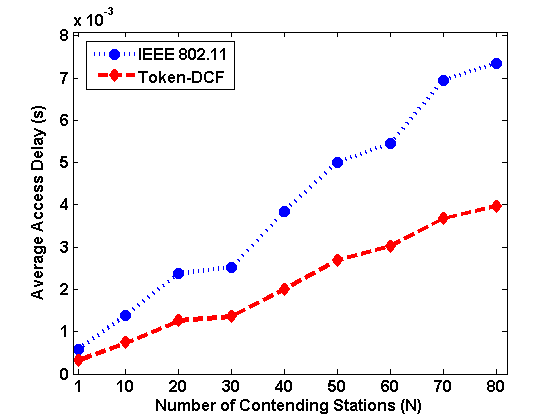}
\end{center}
\caption{Average access delay (area=1500mx1500m)}\label{delay_A1500_ps1500}
\end{figure}

In this section, we study performance of Token-DCF in multihop wireless networks. We consider two network sizes; $800m$x$800m$ and $1500m$x$1500m$. Recall that the effective transmission range in the simulations is limited to $250$ meters and carrier sense range is limited to $550$ meters. Traffic is full buffer CBR and all flows are single hop. The payload size is 1500 bytes. System throughput and access delay of the networks with size $800m$x$800m$ versus number of contending stations are presented in Figures \ref{thr_A800_ps1500} and \ref{delay_A800_ps1500}, respectively. Comparing Token-DCF and 802.11 DCF in these two figures, we can see that throughput gain is a factor of $1.8-2$ and access delay is reduced by a factor of $0.53-0.58$. For the networks of size $1500m$x$1500m$, system throughput and access delay are presented in Figures \ref{thr_A1500_ps1500} and \ref{delay_A1500_ps1500}, respectively. In this case, throughput gain is a factor of $1.9$ and access delay is reduced by a factor of $0.52-0.55$.

Considering Figures \ref{thr_A150_ps1500}-\ref{delay_A1500_ps1500}, we see that similar performance improvement is obtained by Token-DCF in single hop and multihop networks. In multi-hop networks, Token-DCF improves the channel utilization in each transmission range.
\section{Stations with unsaturated traffic}
\begin{figure}[t]
\begin{center}
\includegraphics [width= 96 mm]{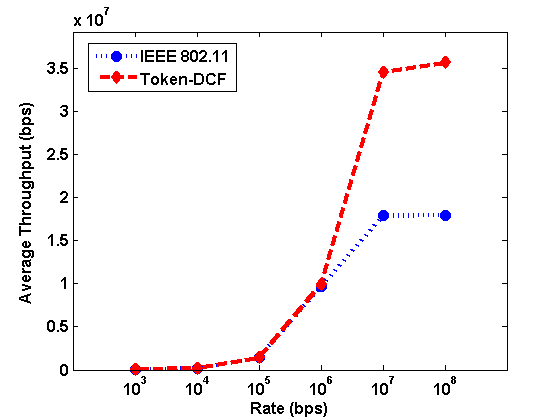}
\end{center}
\caption{System throughput (Pareto traffic)}\label{thr_pareto}
\end{figure}

\begin{figure}[t]
\begin{center}
\includegraphics [width= 96 mm]{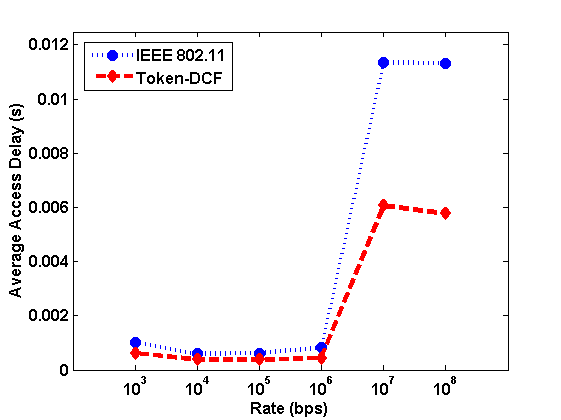}
\end{center}
\caption{Average access delay (Pareto traffic)}\label{delay_pareto}
\end{figure}

Having shown the performance improvement of Token-DCF over 802.11 for saturated networks, we further identify its performance in networks that have less traffic load. The purpose of this set of simulations focuses on comparing the performance of Token-DCF with 802.11 when varying the traffic load from low to high. On/Off traffic with burst times and idle times taken from pareto distributions is used. Configuration parameters are as follows; Packet size is 1500 bytes. Average on time for generator is $50ms$. Average off time for generator is also $50ms$. We perform simulations for randomly generated networks of size $150m$x$150m$. There are a total of $20$ one hop flows. Each source station generates its packets independently and the packet arrival rate of each station during on time is Rate. Rate (sending rate during on time) is varied between $10^3$ $bps$ and $10^8$ $bps$. With Rate $=$ $10^3$ $bps$ and 1500 byte packet size, the traffic demand is far
below the network capacity. When gradually varying Rate 
from $10^3$ to $10^8$ bps, offered load is increased
from small to very large. The corresponding aggregate
throughput and average access delay are presented in Figures \ref{thr_pareto} and \ref{delay_pareto}, respectively. When the network load is
very low, station queues are empty most of the time in which case no station is chosen as the privileged station. Under low load, Token-DCF behaves very similar to 802.11 DCF. Their performance starts to diverge when the network is loaded more heavily. The saturation
throughput of Token-DCF is approximately $2$ times of 802.11.

\section{Networks with TCP traffic}
\begin{figure}[t]
\begin{center}
\includegraphics [width= 96 mm]{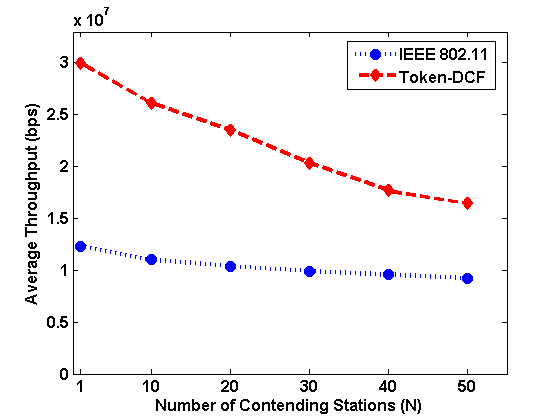}
\end{center}
\caption{System throughput (TCP traffic, area=150mx150m)}\label{thr_A150_ps1500_TCP}
\end{figure}

\begin{figure}[t]
\begin{center}
\includegraphics [width= 96 mm]{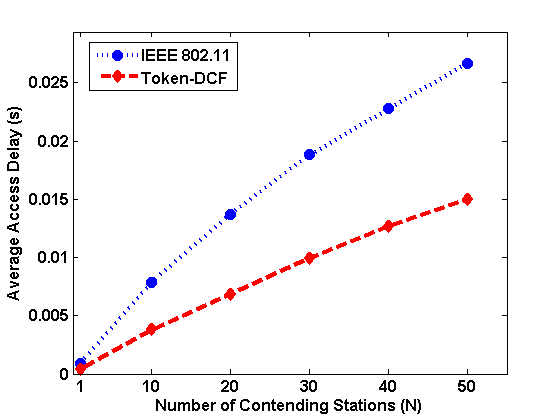}
\end{center}
\caption{Average access delay (TCP traffic, area=150mx150m)}\label{delay_A150_ps1500_TCP}
\end{figure}

In this section, we study Token-DCF's performance under TCP traffic. We perform simulations for networks of different sizes, i.e., $150m$x$150m$, $800m$x$800m$ and $1500m$x$1500m$. Packet payload size is $1500$ bytes. Transmitting and receiving stations of each flow are placed uniformly at random in the area.  As a result, for networks of size $800m$x$800m$ and $1500m$x$1500m$, where network is multi-hop, flows are also multi-hop. DSDV (Destination-Sequenced Distance-Vector Routing) is used as the routing protocol. Figures \ref{thr_A150_ps1500_TCP} and \ref{delay_A150_ps1500_TCP} show the total throughput and average access delay for single-hop networks of size $150m$x$150m$. As presented in these figures, throughput gain is a factor of $1.8-2.4$ and access delay is reduced
by a factor of $0.42-0.56$. Comparing Figures \ref{thr_A150_ps1500} and \ref{thr_A150_ps1500_TCP}, we can see that the performance improvement of Token-DCF over 802.11 is similar for both TCP and saturated CBR traffic. When traffic is TCP, although buffer of stations might not be fully backlogged, stations might have few packets backlogged in their transmission queue, in which case privilege can be given to one of the stations. This results in decreasing the idle time and increasing the throughput. 

\begin{figure}[t]
\begin{center}
\includegraphics [width= 96 mm]{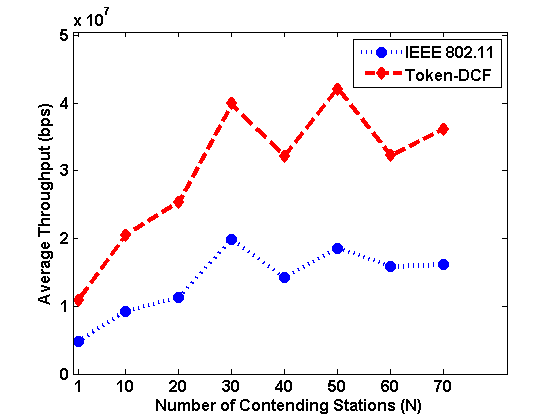}
\end{center}
\caption{System throughput (TCP traffic, area=800mx800m)}\label{thr_A800_ps1500_TCP_multiHopFlow}
\end{figure}

\begin{figure}[t]
\begin{center}
\includegraphics [width= 96 mm]{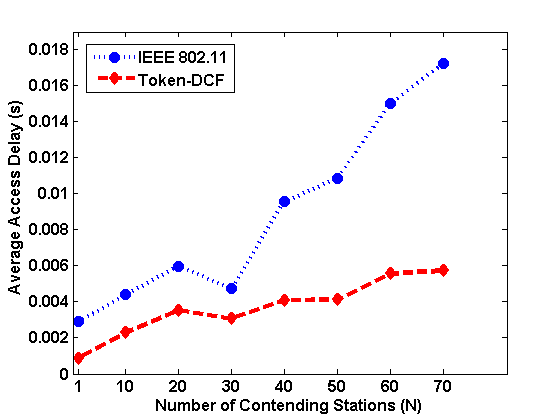}
\end{center}
\caption{Average access delay (TCP traffic, area=800mx800m)}\label{delay_A800_ps1500_TCP_multiHopFlow}
\end{figure}

\begin{figure}[t]
\begin{center}
\includegraphics [width= 96 mm]{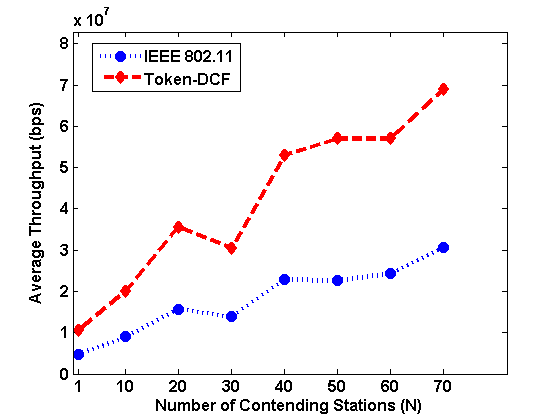}
\end{center}
\caption{System throughput (TCP traffic, area=1500mx1500m)}\label{thr_A1500_ps1500_TCP_multiHopFlow}
\end{figure}

\begin{figure}[t]
\begin{center}
\includegraphics [width= 96 mm]{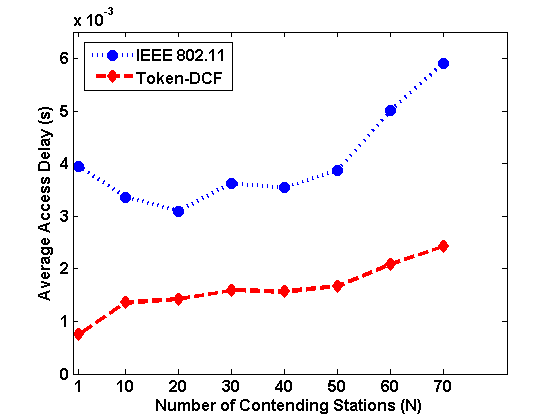}
\end{center}
\caption{Average access delay (TCP traffic, area=1500mx1500m)}\label{delay_A1500_ps1500_TCP_multiHopFlow}
\end{figure}

The results for network size of $800m$x$800m$ are shown in Figures \ref{thr_A800_ps1500_TCP_multiHopFlow} and \ref{delay_A800_ps1500_TCP_multiHopFlow}. In these networks, throughput gain is a factor of $2-2.3$ and access delay is reduced
by a factor of $0.3-0.65$. Figures \ref{thr_A1500_ps1500_TCP_multiHopFlow} and \ref{delay_A1500_ps1500_TCP_multiHopFlow} present total throughput and average access delay for networks of size $1500m$x$1500m$. Throughput gain is a factor of $2.2-2.5$ and access delay is reduced
by a factor of $0.18-0.45$. Although flows are multi-hop in these networks, since Token-DCF improves
the channel utilization in each transmission range, total throughput gain and delay reduction is similar to single-hop networks. 
\section{Conclusion}\label{conclusion}
This paper presents the design and performance evaluation of Token-DCF. Token-DCF is a distributed media access protocol that uses an overhearing mechanism to schedule network stations for transmission on the wireless medium in an efficient manner. The design goal of Token-DCF is to reduce both idle time and collision time. Our simulation results show that Token-DCF significantly improves the performance in terms of system throughput and access delay.

\end{document}